# *MBTModelGenerator*: Automated Reverse Engineering of Test Models from Clickstream Data for Model-Based Testing of Web Applications

| Vahid Garousi<br>Queen's University Belfast, UK<br>Azerbaijan Technical University, Azerbaijan<br>v.garousi@qub.ac.uk | Alper Buğra Keleş, Yunus Balaman<br>Testinium A.Ş., Istanbul, Türkiye<br>{alper.keles, yunus.balaman}@testinium.com |
|---|---|
| Sasidhar Matta<br>Queen's University Belfast, UK<br>smatta01@qub.ac.uk | Zafar Jafarov, Aytan Mövsümova, Atif Namazov<br>Azerbaijan Technical University, Azerbaijan<br>{zafar.cafarov, aytan.movsumova,<br>atif.namazov}@aztu.edu.az |

**Abstract:**

**Context:** Model-Based Testing (MBT) was first introduced in 1970's, and has the potential to improve efficiency and effectiveness of testing. However, its adoption—especially for web applications—has been hindered by the effort required to manually design MBT models, and keep them updated.

**Objective:** Based on the above challenge in a real industrial context, this study introduces an automated approach to reduce that effort by reverse engineering MBT models from clickstream data captured during users' interaction with web applications.

**Method:** We have developed and present in this paper an open-source tool, named *MBTModelGenerator*, which logs user interactions via a lightweight JavaScript module in the front-end, and transmits them to a REST API backend. These interactions are then transformed into directly executable MBT models in the input format of an open-source MBT tool named *GraphWalker*.

**Results:** The tool was evaluated on two representative open-source web applications, Spring PetClinic and a Task Manager web app, and is under evaluation in several large-scale industrial testing projects. The generated MBT models accurately reflected user navigation flows and could be executed in the GraphWalker MBT tool without any manual changes. Using the tool has significantly reduced the effort of MBT model design by more than 90%, while still allowing test engineers to inspect and refine the generated models for completeness.

**Conclusion:** Our approach facilitates lightweight adoption of MBT by automating model generation, which is the most effort intensive phase of MBT. To ensure correctness and completeness, the generated models should still be reviewed by test engineers —but that effort remains substantially lower than designing MBT models from scratch. The tool is in active industrial use and available as open-source for reuse and further development.

# 1 INTRODUCTION

## 1.1 Context and motivations

Systematic and adequate testing of software systems is a costly activity, but so do the costs caused by software defects due to inadequate testing [1]. In a quest to increase effectiveness and efficiency of testing, software engineers have used test automation [2] since early 1990's. While most practitioners use automation for the test execution phase, test automation is "*not just for test execution*" [3], i.e., it can be used in other test activities such as test-case design.

Model-Based Testing (MBT) [4, 5] is a black-box testing approach that enables automated test generation (test-case design) from certain types of behavioral models. MBT can offer significant benefits by enabling the systematic generation of large sets of test sequences (paths) through automated traversal of behavioral test models (such as state-charts and activity diagrams) that represent the behavior of the system under test (SUT). Instead of manually writing individual test cases, test engineers can define a behavioral model—typically composed of states (vertices) and transitions (edges)—and then use test generation algorithms to automatically produce diverse test paths that satisfy specific coverage criteria (e.g., all states, all transitions, or specific path lengths). As a result, MBT can enhance test coverage, ensures consistency with expected behavior, and supports automation at scale, making it particularly valuable for complex and rapidly evolving systems. When integrated with execution tools such as Selenium (selenium.dev) or Appium (appium.io), MBT can provide end-to-end automation of web and mobile applications, boosting both efficiency and effectiveness of testing [6-9].

Our industrial setting is Testinium A.Ş., a large software testing company with 700+ test engineers, headquartered in Istanbul, Türkiye. At Testinium, we have actively used MBT since 2019 to test a wide range of client applications (web and mobile apps). Through a systematic tool evaluation process reported in an earlier paper [7], we have adopted the open-source MBT tool GraphWalker (graphwalker.github.io), which has served our projects well. Our MBT experience have been shared in previous studies [7].

## 1.2 Problem statement

Although MBT has existed for over 55 years now—starting with a technical report by IBM [10] in 1970 presenting a MBT test tool to generate test cases based on Cause-Effect Graphs—, "*most developers [still] don't view MBT as a mainstream [testing] approach*" [11]. Numerous studies have mentioned key barriers to industrial MBT adoption [5, 11-16], including: (1) lack of awareness and training about MBT, (2) issues with usability of existing MBT tools, (3) non-trivial upfront modeling effort, and (4) resistance from stakeholders unwilling to invest in MBT-specific training and the upfront modeling effort.

Most MBT success stories come from domains such as aerospace and embedded systems [17], while adoption in enterprise-scale web and mobile applications has lagged. Despite its potential benefits, MBT adoption in practice is hindered by the non-trivial effort required to manually construct (design) the models to be fed into MBT tools. As a white-paper [18] by Atos, a large international software consulting firm, notes: "*Modeling [of MBT models] can require considerable effort.*"

In our industrial context (Testinium A.Ş.), where numerous test automation projects run in parallel, we have also observed the up-front modeling effort (cost) of MBT has been a major issue [7]. The need to train test engineers on MBT, and to allocate skilled test engineers for model design, combined with the time pressure of fast-paced delivery and test cycles, often rendered manual MBT model design impractical. As a result, teams tend to rely on conventional manual test-case design and then script-based automation approaches, even when MBT could provide higher coverage and long-term maintainability [7]. We have thus observed that overcoming this modeling bottleneck (up-front modeling effort) is critical for enabling broader adoption of MBT in our context, and in the wider industry in general.

Due to the above practical barriers, we beleive lightweight, pragmatic MBT approaches are needed to address this gap [16]. These are the issues that the work reported in this paper aims to address.

## 1.3 Reverse engineering as a practical enabler for MBT

Justifying the initial effort investment for designing any software models (e.g., using the UML modeling language) is a well-known industrial challenge in Model-Driven Software Engineering (MDSE) [19-23]. A widely used strategy to address this challenge has been (often automated) reverse engineering of models—such as UML diagrams—from either static software artifacts (e.g., source code) or dynamic execution data (e.g., traces or logs) [24-29]. In testing, this concept has led to tools for GUI ripping [30], and dynamic crawling [31-34].

Given that model creation is particularly time-consuming in MBT, we sought a readily available source of dynamic interaction data that could be leveraged for model extraction. Inspired by the success of reverse engineering from execution traces, we identified clickstream data [35, 36] as a promising input candidate. Clickstream data are byproducts of end-users' interactions with web applications. Clickstream data (also called click-path) captures sequences of hyperlinks and UI events in the order they were viewed or activated by users [35, 36]. Clickstream data are widely used in analytics tools such as Google Analytics, in which they are often called behavior flow (of users) [36]. Thus, we use reverse engineering from clickstream data as our approach in this paper, which is a promising path for reducing the modeling burden in MBT.

Motivated by these insights, we developed a tool-supported approach that automatically reverse engineers MBT models from web applications' clickstream data. The tool, named *MBTModelGenerator*, transforms raw clickstream logs into MBT models compatible with the GraphWalker tool, which has been the adopted MBT tool in our industrial setting since 2019 [7]. Our approach aims to support immediate test generation and execution—substantially decreasing the manual modeling effort, and thus lowering the barrier to MBT adoption in practice. We shall note that, to ensure correctness and completeness, the generated models should still be reviewed by test engineers —but that effort remains substantially lower than designing MBT models from scratch (empirical data about these will be provided in Section 5).

### 1.4 Summary of contributions and paper structure

The contributions of this paper are:

1. A practical approach and open-source tool for reverse engineering MBT models from clickstream data, targeting real-world web applications
2. Design and implementation details of the MBTModelGenerator tool, including architecture, API, and event-handling logic
3. A case-based empirical evaluation using two realistic systems under test (SUTs), validating tool's feasibility and illustrating usage in practice

The remainder of this paper is organized as follows. Section 2 provides the background and a review of related work. Section 3 presents our proposed approach for reverse engineering MBT models from clickstream data. Section 4 introduces our supporting tool, MBTModelGenerator, describing its architecture, design, and implementation. Section 5 reports on the empirical evaluation of the tool using two real-world case-study systems. Section 6 discusses practical insights, limitations, and implications for industrial adoption. Section 7 concludes the paper and outlines directions for future work.

## 2 BACKGROUND AND RELATED WORK

This section begins with an overview of MBT and its adoption challenges in industrial settings. We then review techniques for reverse engineering test models, introduce the GraphWalker tool used to design and execute our MBT models, and summarize the industrial context in which our work was conducted.

### 2.1 Overview of MBT, its history, and challenges of its adoption in practice

(BT) has been studied and practiced for over five decades, beginning with IBM's TELDAP tool in 1970, which used cause-effect graphs (as a type of models) for test generation [10]. Since then, a wide range of MBT approaches and tools have emerged [37-39], supporting different model semantics (e.g., UML, BPMN), different levels of abstraction, different execution modes (offline vs. online), and different test selection criteria (e.g., model coverage or fault-based strategies) [17, 37, 40].

Despite its benefits, MBT adoption has remained limited in enterprise software testing. While domains such as aerospace, embedded systems, and telecommunications have seen a broader uptake of MBT, e.g., [12, 41-46], usage of MBT in enterprise-grade web and mobile application testing has lagged. A key reason is that "*most developers [still] don't view MBT as a mainstream [testing] approach*" [11], with barriers including: steep learning curves, tool complexity, lack of training, and significant modeling effort [5, 11-16].

### 2.2 Overview of an open-source MBT tool: GraphWalker, in use in our industry context

The MBT models in our approach are designed for and executed using *GraphWalker* (graphwalker.github.io), an open-source MBT tool that supports state-based model execution for automated testing. As discussed in Section 1, through a systematic tool evaluation process [7], we adopted this tool for our MBT practices in year 2019, and have used it since then in our industrial setting [7]. We thus provide a brief overview of the tool, since the reverse engineering approach and tool reported in this paper is in the context of this MBT tool.

GraphWalker uses finite-state machine (FSM)-like models, composed of vertices (states/pages) and edges (transitions/actions). Test cases are generated dynamically based on traversal algorithms such as random walk, edge-coverage, or A* pathfinding. GraphWalker has been used in both academic and industrial settings for MBT of web and mobile applications [47-49].

GraphWalker also integrates with Selenium-based test automation and supports embedding test logic directly into edge or vertex annotations. Its flexibility and support for JSON-format models has made it the ideal tool for our approach and context, as reported in [7]. Therefore, MBTModelGenerator is designed to produce output model files, fully compatible with GraphWalker's model execution engine, allowing immediate use for automated test runs.

It would be informative for the reader at this point to see a concrete example of MBT test artifact in the GraphWalker tool. One of the large MBT projects that we reported in [7] was the MBT project who goal was to test one of the company's own tools (developed in-house). The tool is also called by the company name, Testinium (testinium.io), which is a web-based test management tool, based on the Selenium framework. We show in Figure 1 the Login page and the main Dashboard page of this example SUT. Upon a successful login by the user, the system would show the Dashboard page.

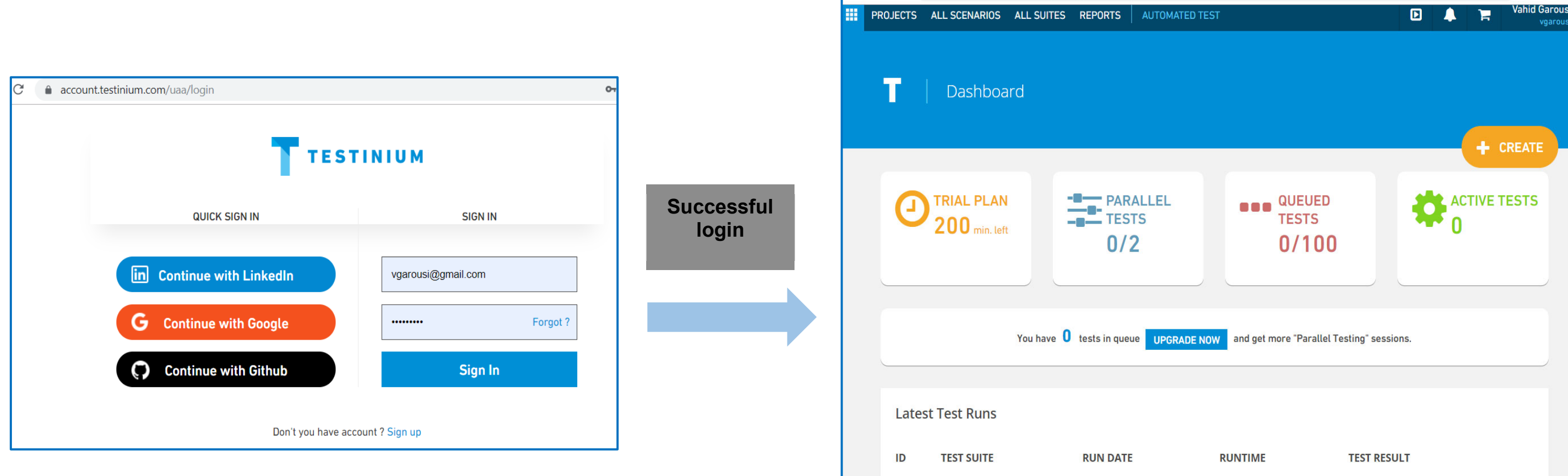

**Figure 1- Login feature of the example SUT: Testinium (testinium.io) [7]**

A MBT best-practice when using the GraphWalker tool is to follow modular design, and to design a separate MBT model for each web page of a given SUT [5, 7]. We show in Figure 2 one of the 18 MBT test models of a large MBT test suite for testing the *login* page [7]. Names of the other 17 test models of this SUT can be seen in the "tabs" in top of the web-based test tool view in Figure 2, and they can be accessed from those tabs.

This *Login* test model verifies the various valid and invalid login scenarios supported by Testinium, including: (1) Standard login with correct and incorrect credentials, (2) Third-party logins (Google, LinkedIn, GitHub), (3) Forgot password flows for registered and unregistered users, and (4) Input validation (e.g., empty fields, invalid email formats).

Each state (node), labeled as `v_*`, represents a user-visible application screen or condition, e.g., (1) `v_Start`: The starting point of the test sequence, and (2) `v_Verify_In_Login_Page`: Main login interface with multiple options. Test assertions / verifications are developed for each node (more on that aspect below). Transitions (edges) are labeled as `e_*`, and each transition is a user action (e.g., clicking a button), e.g., `e_Click_Signin`.

In the context of the GraphWalker tool, test engineers need to provide the behavior of each node and edge of the MBT model, in Java using the Selenium framework. For example, for the edge `e_valid_login` in Figure 2, we developed the Java test code shown in Code Listing 1. In this example Selenium Java code, to conduct a valid login, the username and password fields are first located. Then, a correct combination of username and password values are entered in those fields. To find the HTML button for the "Sign in", a CSS selector path is given. The sign-in button is finally clicked programmatically.

More details about the MBT practice in GraphWalker and our MBT projects have been reported in [7]. Furthermore, several extensive demo videos can be found online [50].

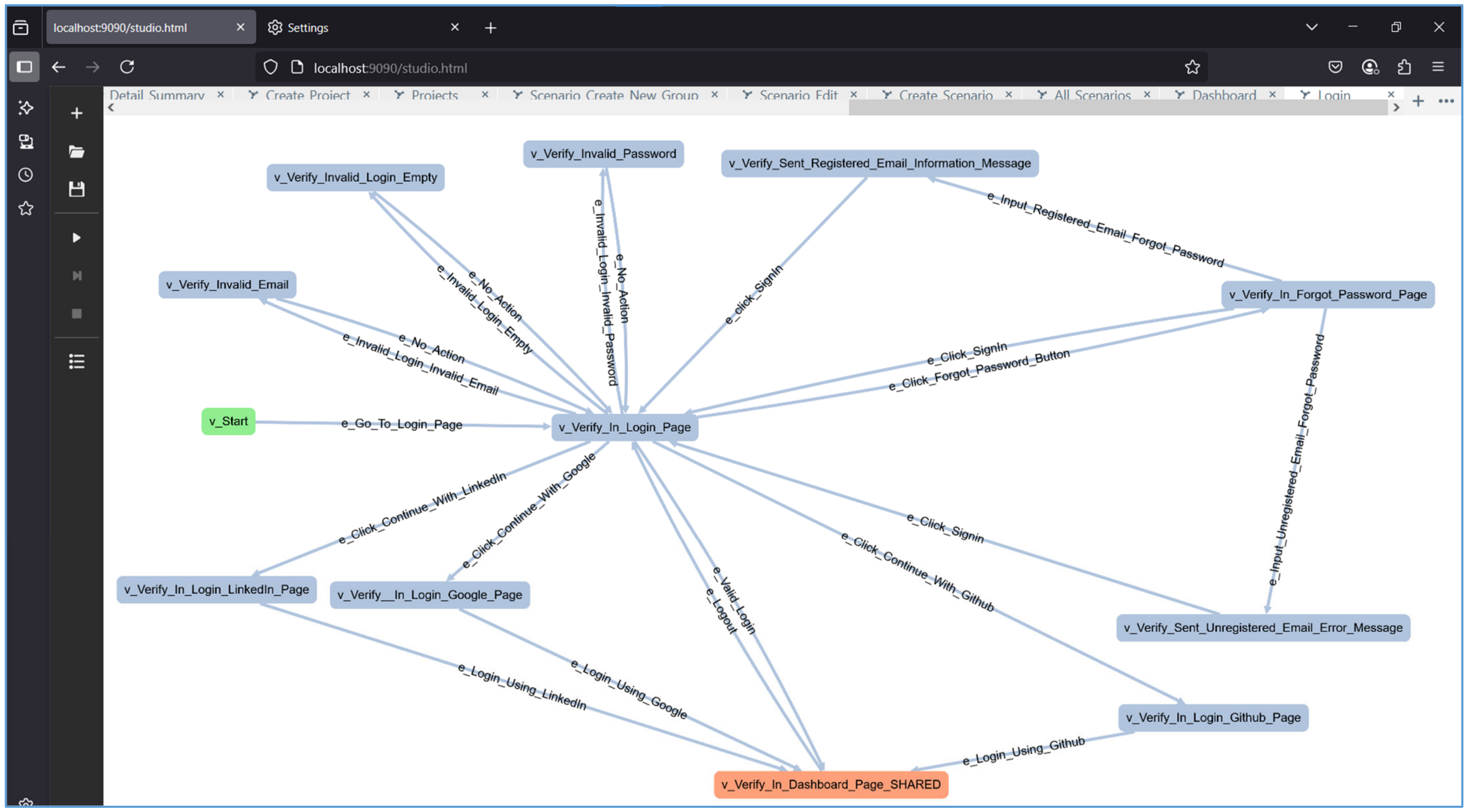

**Figure 2- One of the 18 MBT test models for testing the *login* page of the SUT: Testinium (testinium.io), opened in the test tool: *GraphWalker Studio*. Names of the other 17 test models can be seen in the "tabs" in top of the web-based test tool, and they can be accessed from there**

```
public void e_valid_login() {
   WebElement userNameElement =
       methodsPage.findElement(By.id("username"));
   userNameElement.clear();
   userNameElement.sendKeys(email);
   WebElement passwordElement =
       methodsPage.findElement(By.id("password"));
   passwordElement.clear();
   passwordElement.sendKeys(password);
   methodsPage.findEleminrent(By.cssSelector(
     "input[class$=\"login-page__submit-btn\"][value=\"Sign In\"]")).click();
}
```

**Code Listing 1-Java Selenium code implementing the behavior for edge `e_valid_login` in Figure 2**

### 2.3 Industrial context: MBT at Testinium A.Ş., and the need to make MBT more efficient

The context of this study is Testinium A.Ş., a large software testing service-provider company headquartered in Istanbul, Türkiye. With over 700 engineers and a large customer base in Türkiye and many other countries, the company provides testing services and also offers three flagship test tools: Testinium (testinium.io, online test management tool based on Selenium), Loadium (loadium.com, online load testing tool), and Oobeya (oobeya.io, software quality analytics platform).

MBT has been used pragmatically in the company since 2019 to improve effectiveness and efficiency of test automation [7]. The initiative emerged from an industry-academia R&D collaboration, which was initially supported by a European ITEA4 project, named "TESTOMAT – The Next Level of Test Automation" (testomatproject.eu). Several prior studies have documented this fruitful R&D project [7, 51-54].

In addition to MBT, the company has explored many other ways for improving effectiveness and efficiency of testing, e.g., enhancing BDD practices via the Gauge BDD tool (gauge.org), applying test maturity models (e.g., TMMi [53]) and earning the TMMi Level-3 certificate (tmmi.org/accredited-certifications), improving test result visualizations, and optimizing regression test prioritization. The MBTModelGenerator tool initiative has emerged from this broader commitment to innovation in industrial testing practices.

### 2.4 Related work on reverse engineering of models to be used for testing web application

Reverse engineering has long been used to reduce the modeling burden in software engineering tasks such as design recovery and test generation. In the context of testing web applications, two main types of runtime artifacts have been leveraged for this purpose: (1) dynamically generated execution traces, and (2) clickstream data, which capture actual user interactions with web pages. This section reviews prior efforts in both directions and positions our work in this landscape.

A related stream of work focuses on dynamic crawling techniques, such as those implemented in Crawljax [31, 57] and ProCrawl [34], which dynamically interact with web applications to discover UI states and transitions. These tools systematically explore the DOM to build navigational models and extract execution flows. While they have shown potential in generating test cases or state-based models, they typically simulate user actions in a synthetic or exhaustive manner rather than capturing real-world user flows.

Other efforts have targeted the automated extraction of GUI models or state machines through GUI ripping or dynamic analysis, particularly for desktop and mobile applications [30][31-34]. However, these tools often require intrusive instrumentation or specialized setups, which limits their use in modern web environments, especially in production.

Clickstream data—also known as click-paths—are structured sequences of user navigation and UI interaction events collected during web application usage [36, 55]. These data sources are widely used in web analytics platforms, such as Google Analytics' "Behavior Flow" feature, to study user behavior patterns. More recently, researchers and practitioners have begun exploring clickstream data for testing purposes, including navigation modeling, reliability evaluation, and user-centric coverage analysis [35, 56].

Despite these advances, few existing tools generate executable MBT models that are directly compatible with established MBT tools such as GraphWalker. Moreover, many approaches rely on synthetically generated traces, which may not reflect realistic user behavior or common usage patterns. Our approach addresses both of these gaps: we leverage *actual* user interactions via clickstream logs—thus capturing realistic paths—and produce directly executable MBT models usable in GraphWalker, enabling immediate automation of testing workflows without manual transformation or post-processing.

In summary, while prior tools and studies have explored aspects of automated model generation for testing, our work is distinguished by its (1) use of real-world clickstream data, (2) lightweight implementation without requiring intrusive instrumentation, and (3) support for end-to-end MBT workflows through model generation in GraphWalker format. These aspects make our solution particularly suited for pragmatic, industrial-scale testing scenarios where rapid model generation and automation are critical.

## 3 Proposed Approach: Reverse engineering MBT models from clickstream data

We present in this section the conceptual design and rationale behind the approach, including how clickstream events are interpreted and structured as MBT artifacts. Detailed implementation aspects of the supporting tool—including software architecture, APIs, and reverse engineering algorithm—will be discussed in Section 4.

### 3.1 Core idea and rationale

Building on the reverse engineering techniques discussed in Section 2, our proposed solution uses clickstream data—naturally occurring UI event traces from user interaction with web applications—to automatically generate MBT models. By converting the clickstream logs into GraphWalker-compatible models, we enable test teams to bypass manual modeling while still generating executable test artifacts. This approach is especially suited for agile and regression testing contexts where quick feedback and model updates are needed.

### 3.2 Clickstream data: Definition and characteristics

Clickstream data refers to the sequence of user interactions with a web application, including clicks, page loads, form submissions, and other UI events. Each event typically contains metadata such as:

- Event type (e.g., click, input, submit)
- DOM details of the page element that is interacted (e.g., ID, class, tag)
- Timestamp
- URL or API route of the page where the event occurred

This data represents real user navigation paths, often structured as directed acyclic graphs or linear traces. In our approach, we capture this data via a lightweight JavaScript module embedded into the target web application, enabling fine-grained observation of user interactions. Implementation details of that JavaScript module will be discussed in Section 4.4.2.

### 3.3 Assumptions and known limitations of the approach

Our approach is designed with several assumptions to ensure the feasibility and reliability of reverse engineering MBT models from clickstream data:

- The system under test (SUT) is a modern web application accessible through standard desktop browsers.
- User interactions are primarily composed of standard click, input, and navigation events that can be captured using JavaScript instrumentation.
- UI elements involved in interactions can be uniquely identified using attributes such as ID, class, or XPath Selenium selectors[1].
- Each interaction in the clickstream represents a meaningful application behavior, contributing to the underlying test logic.

Given these assumptions, the approach is tailored for web applications that follow predictable, state-based navigation flows. However, there are important limitations to consider:

- Complex interaction types, such as drag-and-drop actions, touch gestures, or multi-tab workflows, are not currently supported. Events triggered by background JavaScript (e.g., AJAX calls or WebSockets) may not be explicitly captured unless tied to visible DOM changes.
- The clickstream-based models are derived only from observed user behavior. Thus, the resulting MBT models represent partial behavioral coverage, based on either manually performed exploratory testing sessions or recorded traces of real usage of the SUT by real end users. As a result, they do not guarantee full completeness.
- Redundant or semantically equivalent states may appear in the generated models, especially in dynamic UIs where different URLs or DOM paths lead to similar views. These may require manual post-processing to consolidate or optimize the model structure.
- Lastly, the generated models may not represent negative scenarios unless such behaviors are explicitly exercised during the session (e.g., entering invalid credentials, skipping required fields).

Despite these limitations, the approach offers a lightweight and low-barrier path to adopting MBT in web testing workflows. It is especially useful for rapidly bootstrapping MBT models during regression testing or continuous integration cycles, where quick model generation is more critical than complete behavioral coverage.

### 3.4 Online vs. offline clickstream capture

Our approach supports both online (real-time) and offline (log-based) modes for capturing clickstream data, offering flexibility across different testing contexts.

In online mode, user interactions are captured in real time as they occur during manual or exploratory testing sessions. Click and input events are logged and processed immediately, allowing dynamic model construction as the tester interacts with the SUT.

In offline mode, pre-recorded clickstream logs are used as input, which are collected from the field after real users have used the system, or are created by exploratory manual testing. These logs typically include structured sequences of events along with contextual metadata (e.g., timestamps, URLs, and element identifiers).

Both modes aim to capture realistic navigation paths and behavioral flows, enabling automated model generation without requiring manual diagramming or pre-defined test scenarios. This dual-mode capability would support a wide range of MBT use cases, from exploratory regression testing to post-production usage analysis.

### 3.5 Mapping clickstream events to MBT models (vertices and edges)

Once collected, the clickstream events are mapped into MBT models using a state-transition structure. In this mapping, based on the GraphWalker modeling semantics (as reviewed in Section 2.2), vertices represent UI states or application views (e.g., login page, dashboard), and edges represent user-triggered transitions (e.g., clicking a button or submitting a form). In other words, each UI state (e.g.., being in a given web page) is modeled as a vertex, and each navigational action (e.g., a button click) is captured as an edge.

---

[1] selenium.dev/documentation/webdriver/elements/locators

Each event in the clickstream sequence—such as a navigation, click, or form interaction—contributes to the construction of the resulting MBT model. A new vertex is created when the user navigates to a previously unseen UI state, while an edge is added for each transition between those states.

This process captures user behavior as executable models. It accommodates repeated transitions and alternate paths, supporting both positive and negative test case generation. While the models are automatically constructed, they can optionally be refined by test engineers to remove redundant nodes, merge semantically equivalent states, or enhance readability. The output models conform to the structure accepted by GraphWalker, enabling immediate integration into automated MBT workflows without the need for further manual changes or transformation.

## 4 Tool Support: MBTModelGenerator

This section presents the implementation details of MBTModelGenerator, covering its architecture, components, and operational flow. While Section 3 focused on the conceptual design of our approach, here we describe the actual engineering of the tool, including the backend and frontend components, REST API design, model generation process, and visualization features.

### 4.1 Tool overview and open-source availability

To operationalize the proposed approach described in Section 3, we developed a supporting tool named MBTModelGenerator. It supports both online (real-time) and offline (log-based) clickstream processing, enabling flexible use across exploratory testing, production monitoring, and CI environments.

MBTModelGenerator is released as open-source software and is publicly available via GitHub: github.com/vgeruslu/MBTModelGenerator. Availability of the tool will facilitate adoption, customization, and extension of the tool by the broader research and practitioner community. Furthermore, a demo video of the tool can be found in youtu.be/vEcFvZEkyb0.

### 4.2 Architecture, design, and usage context of the tool

Figure 3 illustrates the high-level component architecture and operational flow of MBTModelGenerator, for MBT of web applications. The overall process is organized into two stages, as discussed next.

#### 4.2.1 Stage 1: Reverse Engineering MBT models from clickstream data

In this stage, users (typically test engineers or developers) interact with the SUT via a standard web browser. The `GraphWalker.js` module—embedded into the SUT's front-end—monitors browser events such as button clicks, form submissions, and input changes.

These monitored UI events are transmitted as clickstream messages to a Spring Boot-based REST API server that hosts MBTModelGenerator's backend logic. The backend API server is responsible for: initializing model sessions, accepting and processing event data (vertices and edges), tracking session activity via periodic keep-alive signals, and finally saving (exporting) the MBT model into a GraphWalker-compatible JSON file upon session timeout or manual stop.

The client-side (browser) and server-side (REST API) components may be hosted on the same physical machine or distributed across separate nodes, depending on the test environment. This decoupled architecture supports a wide range of deployment configurations and does not require modifications to the SUT itself—only inclusion of the JavaScript monitor (`GraphWalker.js` module) in front-end HTML code is enough for that purpose.

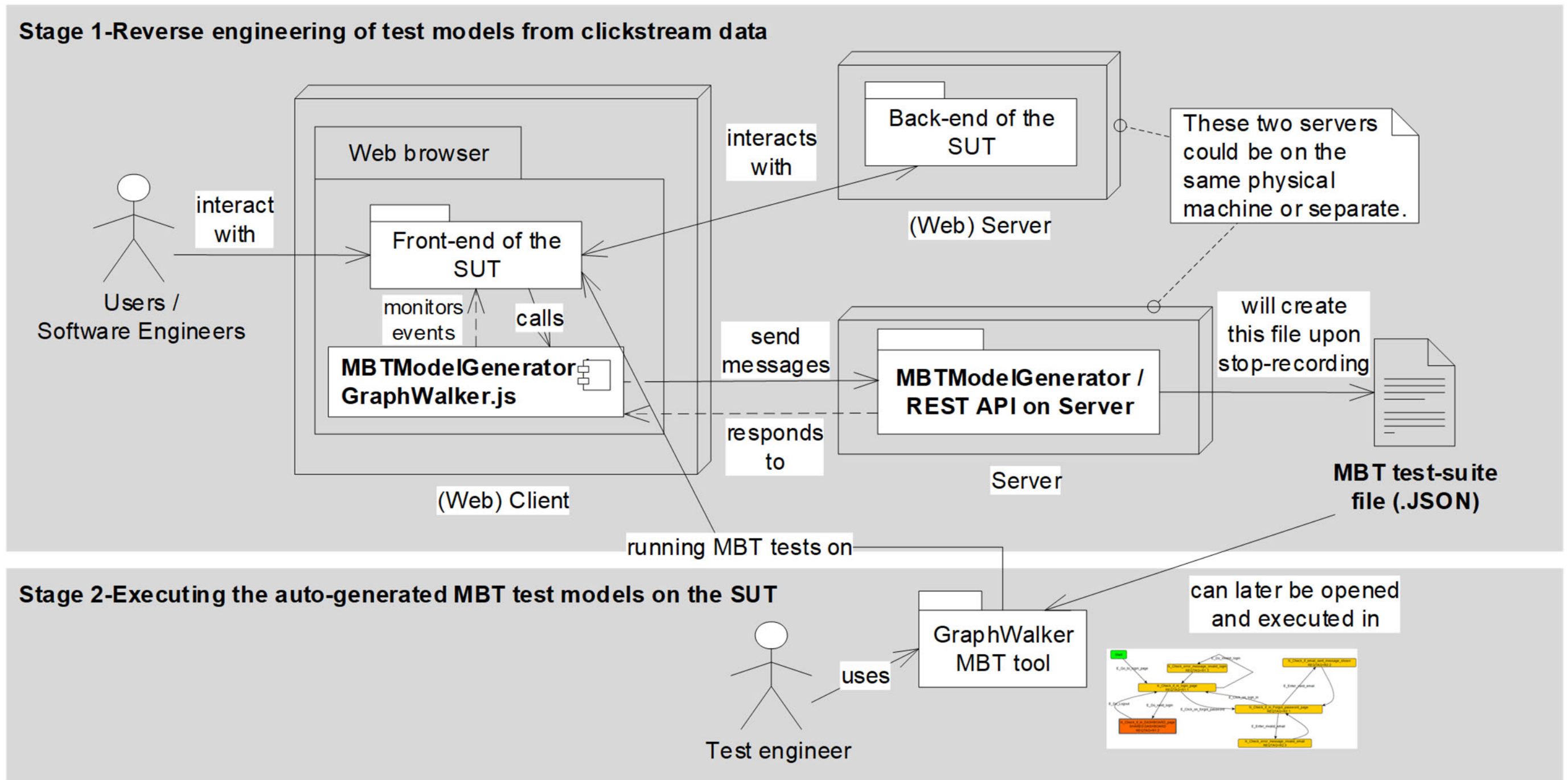

**Figure 3-Architecture diagram of MBTModelGenerator and its usage context**

In addition to the architectural workflow shown in Figure 3, we also present a selected view from the internal code structure of MBTModelGenerator in the form of a dependency diagram. We have done this code reverse engineering using the *Understand* tool (scitools.com), which generates dependency diagrams to support understandability, maintainability and extensibility. Figure 4 presents the high-level module dependencies of the tool's backend implementation. This diagram captures class-level relationships and method-level interactions, offering a closer look at how responsibilities are divided and how key components collaborate.

At the entry point, the `gwmaker` package (`gw` for GraphWalker) contains the primary request-handling components. These classes call the `newGraphWalker()` method defined in `GraphWalkerFactory.java`, which is responsible for creating new instances of the `GraphWalker` class. This method encapsulates the initialization logic and delegates the control of the MBT model lifecycle to the `GraphWalker` class.

Inside the `GraphWalker` class, the following methods are central:

- `setModelName()` assigns a unique name to the current session's model.
- `onVertex()` and `onEdge()` are invoked when a user interaction is logged and map the input into vertex or edge objects, respectively.
- `finalizeGraph()` completes the session, triggering coordinate generation and JSON serialization of the full model.
- The GraphWalker class internally relies on several supporting modules:
- `vertexFactory` and `edgeFactory` construct vertex and edge objects using standardized naming and attribute conventions.
- `lastVertex` and `lastEdge` store the most recently added elements during model construction, allowing the tool to maintain the correct sequence and linkage between user actions and application states.
- `pathGenerator` calculates the graphical path layout of edges for visualization.
- `model` holds the actual graph data structure, while `json` modules handle output serialization.

All these components form a cohesive pipeline, wherein each class has a clearly defined role and communicates through well-scoped interfaces. Notably, many interconnections are labeled with low cardinalities (mostly 1 or 2), which demonstrates the system's tight modularity and low coupling. This dependency view highlights the tool’s alignment with good software engineering practices: encapsulation, factory patterns, and separation of concerns—all contributing to maintainability and extensibility.

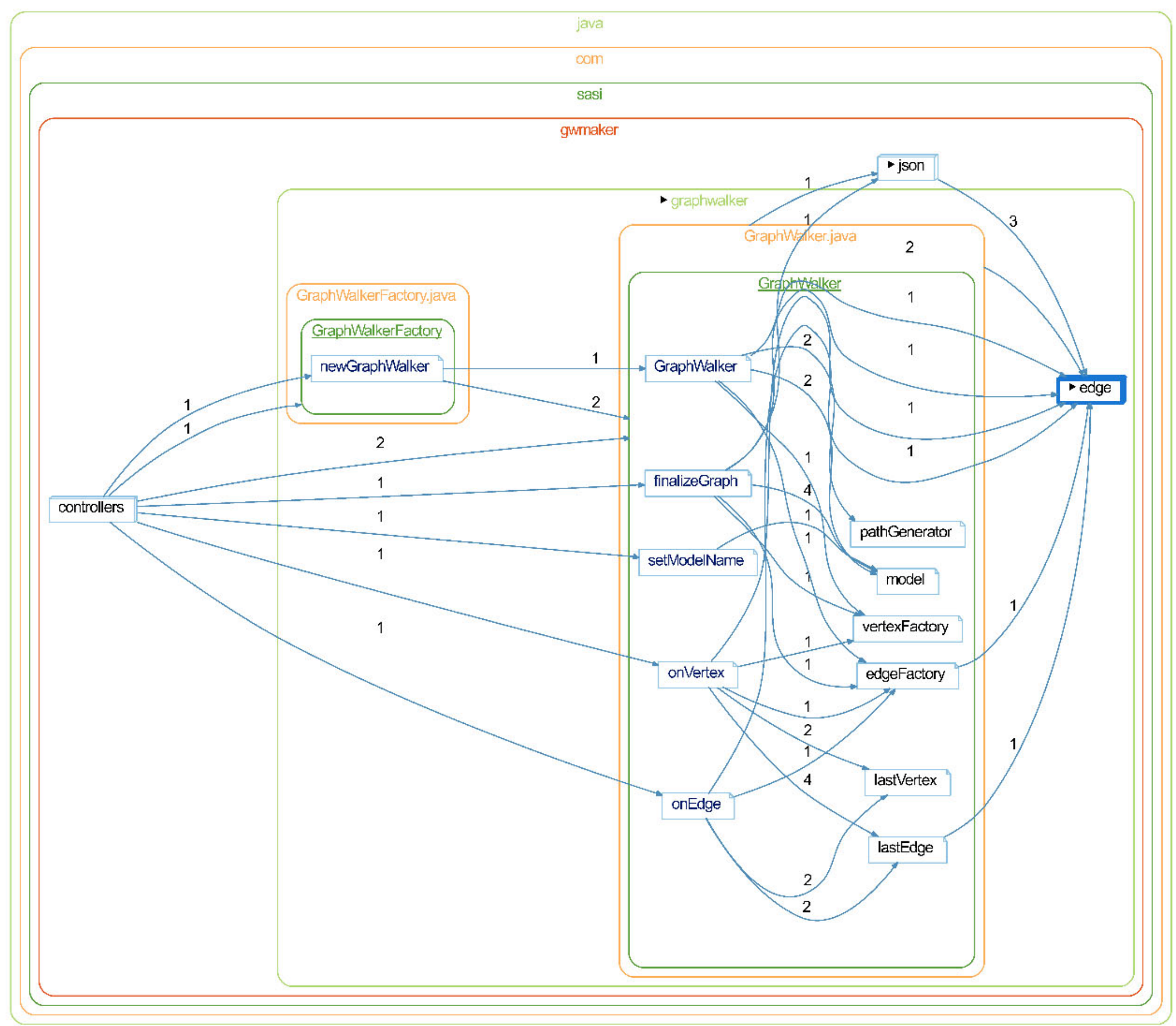

**Figure 4- High-level dependency diagram of the MBTModelGenerator codebase. Generated using the *Understand* tool ([scitools.com](scitools.com))**

Further design details about design and implementation of the MBTModelGenerator tool, including UML design diagrams, and API design, can be found in a technical report [58].

#### 4.2.2 Stage 2: Executing the auto-generated MBT models to test the SUT

Once the MBT test model is generated and saved as a JSON file, test engineers can load it into the GraphWalker MBT tool. GraphWalker reads the model and dynamically generates and executes test paths according to coverage strategies such as random walk or edge-coverage.

The GraphWalker engine interacts with the SUT's front-end through automation scripts (e.g., Selenium WebDriver), executing user actions as described in the model.

This architecture enables a complete end-to-end MBT workflow. It begins with the passive monitoring of user interactions with a web application—either from exploratory testers or actual end-users. These natural interaction traces are then automatically transformed into MBT models without requiring the test engineer to manually design them. The resulting models, saved in a format directly compatible with GraphWalker, serve as executable test artifacts. In the second stage, these models are replayed through GraphWalker, which drives automated test execution on the SUT—ensuring that user-driven behaviors are now systematically tested. In essence, the MBT model acts as a bridge between recorded real-world usage by end users and automated MBT execution.

Having described the overall architecture and internal structure of MBTModelGenerator, we now turn our focus to how the tool components interact at runtime. In particular, the coordination between the browser-based frontend module and the backend server relies on a set of REST API endpoints that govern session initiation, data transmission, and model finalization. The next section details these endpoints and the lifecycle of a typical recording session.

### 4.3 REST API endpoints and session lifecycle

At the core of MBTModelGenerator's runtime operation is the interaction between its frontend JavaScript module (GraphWalker.js) and the Spring Boot-based backend API server. This communication is facilitated through a lightweight set of REST API endpoints. The frontend module, embedded into the web application under test, monitors user interactions in real time and asynchronously transmits those events to the backend, which processes and persists the MBT model.

Each session follows this lifecycle:

- `POST /startrec`
  Once the frontend JavaScript module calls this endpoint, it initiates a new modeling session by providing a unique model name. The backend creates an internal model structure and begins accepting events associated with that session.
- `POST /vertex?name=...`
  Each time the frontend module detects navigation to a new UI state, it calls this endpoint to register a vertex in the model. The backend adds this vertex, labeled with the provided name, representing a unique application screen or state.
- `POST /edge?name=...`
  Whenever a user action such as a button click or form submission is observed, the frontend calls this endpoint to register an edge. This edge links the most recently observed vertex to the next, capturing the transition triggered by that interaction.
- `POST /keepalive`
  While the session is ongoing, the frontend module sends periodic keepalive messages (typically every 3.33 seconds) to ensure the session remains active. Each call resets the backend's session timeout counter.

Session finalization occurs automatically if no keepalive signal is received within a 10-second window. At that point, the backend triggers the `finalizeGraph()` method, which: completes the model structure, computes visual layout coordinates (via PlaneGenerator, details in Section 4.6), and serializes the model into a GraphWalker-compatible JSON file.

This architecture ensures a lightweight, non-intrusive integration into existing web applications. The frontend module does not require any changes to the application's business logic—it simply observes user behavior and streams interaction events to the backend. Together, these REST endpoints enable real-time, automated construction of MBT models that accurately reflect user workflows, ready for immediate test generation and execution.

### 4.4 Implementation details (Spring Boot backend, JS frontend integration)

Based on the system design that we presented in Section 4.2 and 4.3, the implementation of MBTModelGenerator consists of two coordinated components: a backend system developed using the Spring Boot framework, and a frontend JavaScript module, GraphWalker.js, that runs inside the SUT's browser context. We discuss these aspects next.

#### 4.4.1 Spring Boot backend

The backend server provides a set of RESTful endpoints to initiate recording sessions, receive UI events (vertices and edges), and finalize the MBT model. The architecture follows modular design principles, with key responsibilities divided across several classes (as visualized in the high-level and detailed dependency diagrams in Figure 4).

- The `GraphWalkerFactory` class provides the entry point for creating new GraphWalker instances. This abstraction supports session-based test modeling.
- The GraphWalker class orchestrates the core model-building process. Its main methods include:
  - `setModelName()`: assigns a name to the MBT model instance.
  - `onVertex()` and `onEdge()`: process clickstream events into corresponding MBT elements.
  - `finalizeGraph()`: invoked at session end to serialize the complete model into GraphWalker-compatible JSON.
- Supporting classes include:
  - `VertexFactory` and `EdgeFactory`: generate model elements while enforcing naming conventions.
  - `PlaneGenerator`: computes layout coordinates for GraphWalker visualization.
  - `ModelToJsonConverter`: handles model serialization to JSON.
  - `TaskTimer` and `ITaskTimer`: manage session timeouts based on keep-alive signals.
  - `StorageHandler`: writes the final JSON file to disk.

This modular breakdown supports extensibility—for example, by plugging in alternate serialization formats or test model optimizers.

#### 4.4.2 Frontend JavaScript module

The frontend component is implemented in a lightweight JavaScript file (≈170 lines of code) named `GraphWalker.js`, which must be embedded into the web pages of the SUT. It is responsible for:

- Monitoring click, input, and navigation events in the browser DOM,
- Mapping those events to meaningful model elements (e.g., vertices for page states, edges for transitions),
- Asynchronously sending event data to the backend API using HTTP `POST` requests.

The module also initiates and maintains the recording session by:

- Sending a `startrec` request when the test begins,
- Issuing `keepalive` signals every 3.33 seconds to prevent premature session termination,
- Allowing session finalization either automatically (after a 10-second timeout) or manually via a stop call.

Importantly, the module uses the Vanilla JS framework (vanilla-js.com), ensuring compatibility with any DOM-based web application without requiring additional libraries or frameworks. This design enables integration with minimal changes to the SUT, making the approach suitable for lightweight, low-intrusion testing scenarios.

### 4.5 Coordinate generation for GraphWalker visualization

To support graphical model inspection and manual refinement in GraphWalker Studio, MBTModelGenerator automatically assigns 2D layout coordinates to all vertices and edges in the generated MBT model. This is a necessary step because GraphWalker requires positional metadata for rendering the graph visually. Without these coordinates, the model would appear unstructured, and hard to visually understand.

The coordinate generation is handled by the `PlaneGenerator` class, which implements a lightweight layout algorithm. Its primary strategy is to position the vertices evenly around a virtual circle. This circular layout ensures that: vertices are spatially separated; edge crossings are minimized; and the visual structure remains clean and symmetric regardless of the model size.

The approach ensures that the vertices (the rectangles) have an angular separation, they are laid around a circle and this calculation is performed by the `PlaneGenerator` class. This method does not attempt to optimize the layout in terms of graph-theoretic properties (e.g., minimal crossings or edge lengths), but it ensures a consistent and readable visualization that is suitable for quick inspection, debugging, and minor editing.

During the finalization of the model (triggered either by session timeout or manual termination), the following steps are executed:

1. The list of vertices is extracted from the model.
2. Each vertex is assigned a coordinate (x, y) based on angular separation from the center of the layout circle.
3. Edge paths are generated as straight lines connecting the corresponding vertex coordinates.
4. All layout metadata is embedded in the resulting GraphWalker JSON file, in which the entire MBT model is stored.

This coordinate generation step is seamlessly integrated into the tool's pipeline and requires no configuration or tuning from the user. If desired, test engineers can later rearrange the layout manually within GraphWalker Studio for domain-specific readability, but the automatically-generated layout serves as a practical and visually clean default.

## 5 Empirical evaluation

This section outlines the evaluation that we have conducted for assessing the effectiveness and practicality of MBTModelGenerator. Guided by the Goal-Question-Metric (GQM) approach [59], we present the structure of the study, describe the two selected SUTs, and explain how the tool was applied. We then analyze the generated MBT models, assess their correctness and utility, and summarize key findings. The section concludes with a discussion of observed benefits and effort savings observed in the case study.

### 5.1 Goal and research questions

Guided by the GQM approach [59], the goal of the empirical evaluation was to analyze the tool for the purpose of assessing its effectiveness and correctness in automatically generating MBT models from clickstream data, and reducing the manual effort required in designing MBT models from the point of view of test engineers and researchers, in the context of web application testing.

From the above goal, we derived the following research questions (RQs):

- RQ1: Can the tool (MBTModelGenerator) generate MBT models that accurately capture user interactions across different pages and flows of a web application?
- RQ2: Can the generated MBT models be loaded and executed in GraphWalker without any manual corrections or fixes?
- RQ3: How much effort (time) saving would MBTModelGenerator typically provide?

We chose the following metrics (assessment criteria) to answer each of the above two RQs:

- Completeness and correctness of captured interaction flows in the generated MBT model, as compared to the actual user interactions during the exploratory-testing session (RQ1),
- Model loading and execution success in GraphWalker without any manual fixing (RQ2).

### 5.2 Study setup and evaluation criteria

For each SUT, the evaluation followed the following steps:

1. We embedded the JS-based logging module (`GraphWalker.js`) into the SUT's frontend, i.e., one single line in the SUT's HTML calling the JS file.
2. Regarding the extraction of clickstream data, as discussed in Section 3.4, we had two options: either online or offline capture of clickstream data, i.e., recording clickstream data via live usage of each SUT, or getting finding logged usage data of each SUT for an extended period of time (e.g., a week of usage data). For these two chosen SUITs, there was no publicly-available clickstream data, thus our only option was to procced with the online approach for extraction of clickstream data. To do this task as systematically as possible in our time resources, we decided to perform exploratory testing sessions, covering realistic user interactions and workflows with each SUT. This also allowed us to deliberately exercise key features and navigation paths of the SUTs, ensuring that the resulting MBT models were representative and structurally complete. In contrast, even if we had access to offline (passive) logs from real users, they would often reflect incomplete, repetitive, or biased usage patterns, making them less effective for structured evaluation.
3. During the exploratory testing session, we let MBTModelGenerator automatically capture the clickstream data and generate MBT models in real time.
4. We then loaded, the auto-generated MBT model in GraphWalker, and then inspected and validated structural correctness and completeness of the models
5. We checked whether the automatically-generated MBT model could be executed directly in the GraphWalker tool, without requiring any manual correction, syntax fixing, or restructuring

### 5.3 Objects under study (systems under test)

We applied the tool to two representative open-source web applications as SUTs: (1) Spring PetClinic (github.com/spring-projects/spring-petclinic); and (2) Task Manager (github.com/rengreen/task-manager). These were selected for their realistic features and relevance to industry-scale testing contexts.

The Spring PetClinic is an open-source reference web application, developed using the Spring Boot framework. As examples of its UI, Figure 5 shows three example pages. This software is commonly used in many software engineering research papers as a system under analysis, e.g., [60-62]. It includes multiple features, such as: CRUD (Create, Read, Update, and Delete) + search operations on pets and pet owners, assigning veterinarians to pet, and CRUD operations on appointment schedules. These aspects make this SUT a suitable SUT for evaluating whether MBTModelGenerator can handle typical multi-page navigation and form interactions.

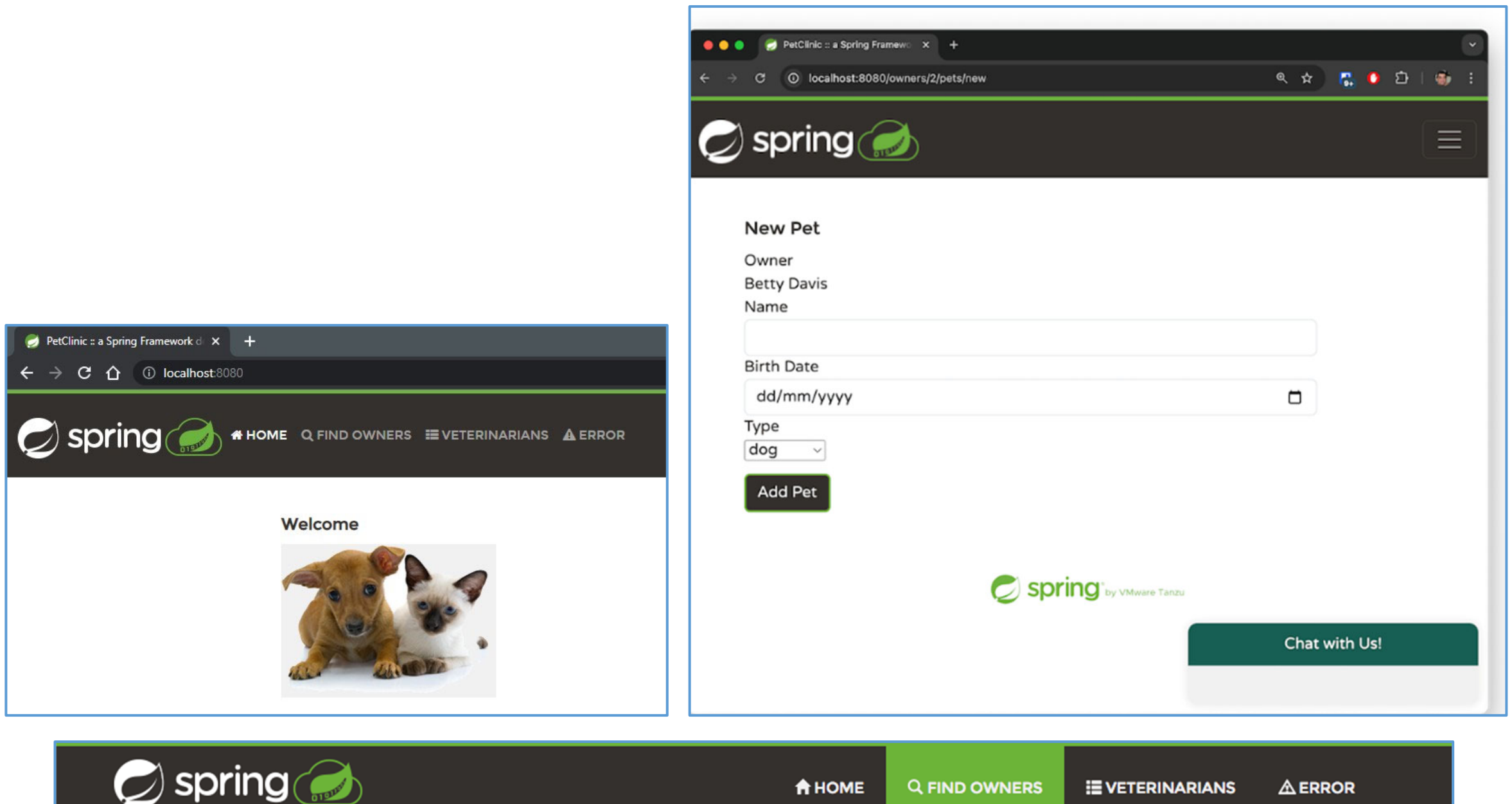

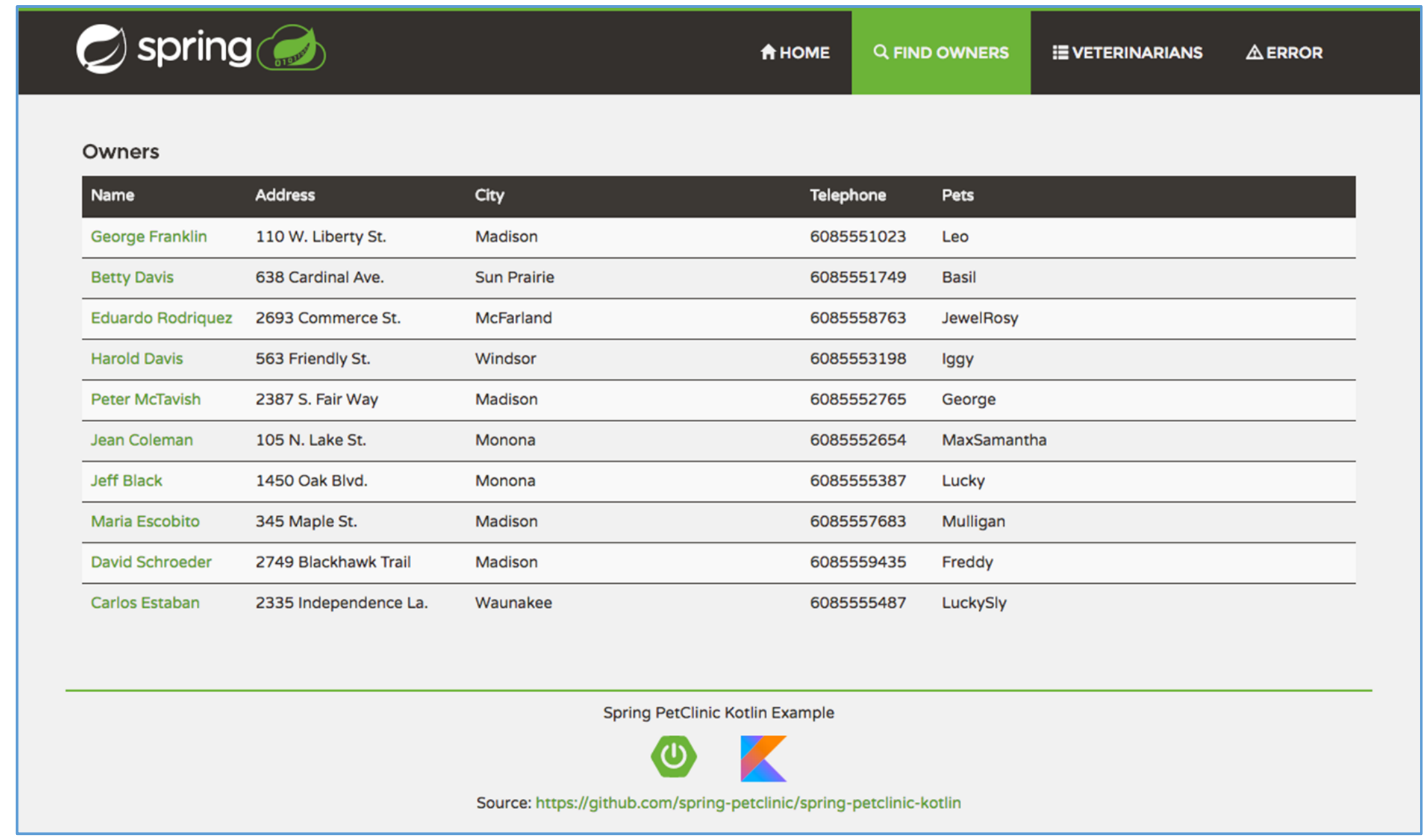

**Figure 5- Three example pages of the Spring PetClinic web app**

The second SUT is a task management web application, called *Task Manager*, which has also been developed using the Spring framework. It supports two actor (user) types: regular users, and managers. Managers can create and assign tasks to users. Regular users can edit and mark their own tasks as complete. As an example UI screenshot, Figure 6 shows the home screen of this web app, after successful login of a regular user.

The application's access control and role-based workflows introduce more complexity compared to the first SUT (PetClinic), including: different pages and UI flows depending on user role, conditional rendering of elements (e.g., admin-only actions), and state transitions that depend on user-specific session context.

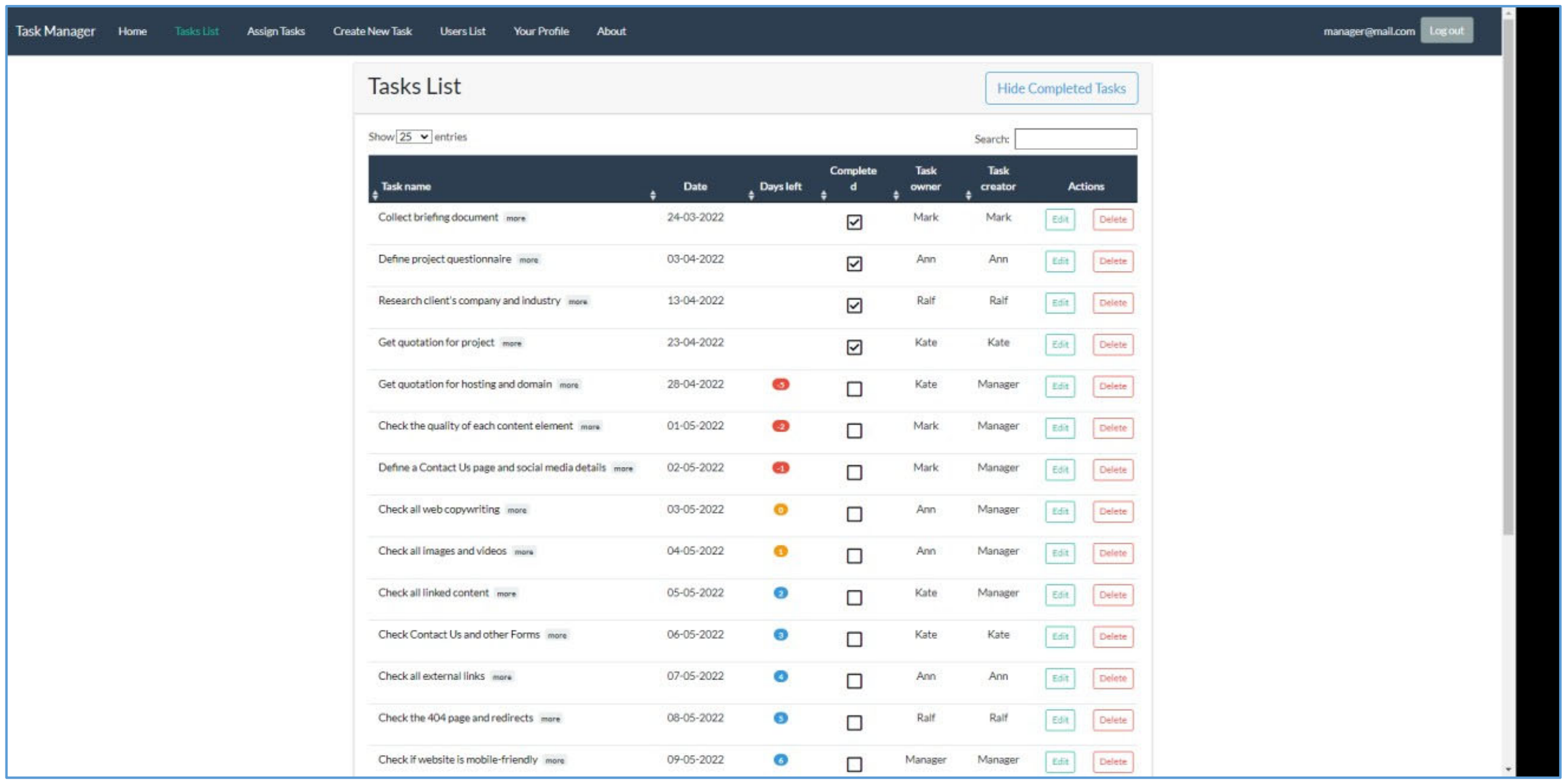

**Figure 6- Home screen of the Task Manager web app**

## 5.4 Applying MBTModelGenerator on each SUT

We report next how we used MBTModelGenerator to reverse engineer MBT models for each SUT.

### 5.4.1 SUT1: Spring PetClinic

We instrumented the PetClinic web app with the tool logging JavaScript module (`GraphWalker.js`). We then conducted exploratory testing for about 20 minutes navigating through different UI paths using, ensuring that main paths are covered. We verified that the clickstream data was collected and processed in real time by the tool. Figure 7 shows the automatically-generated MBT model of this SUT by MBTModelGenerator, loaded and visualized in the GraphWalker tool.

As we can see, the generated MBT model accurately captured states such as the welcome page, owner registration page, and veterinarian list view. For example, when the tester clicked the *Find Owners* button, an edge labeled `e_FindOwners` was logged and connected the vertex `v_WelcomePage` to the `v_FindOwnersPage`.

Upon session finalization (after no keep-alive signal for 10 seconds), the model was serialized (saved) into a GraphWalker-compatible JSON file. The `PlaneGenerator` class automatically positioned vertices with angular separation, producing a clear and symmetric visual layout (as discussed in Section 4.5).

Furthermore, when we pressed the *Run Test* button in GraphWalker (visible in Figure 7), we observed that the automatically generated MBT model was executed without any issues in the GraphWalker tool. Parts of the MBT execution can be seen in the demo video of the tool in [youtu.be/vEcFvZEkyb0](https://youtu.be/vEcFvZEkyb0).

As a result, both empirical assessment criteria for this SUT were passed: (1) Completeness and correctness of the generated MBT model; and (2) Successful loading and execution of the model in GraphWalker.

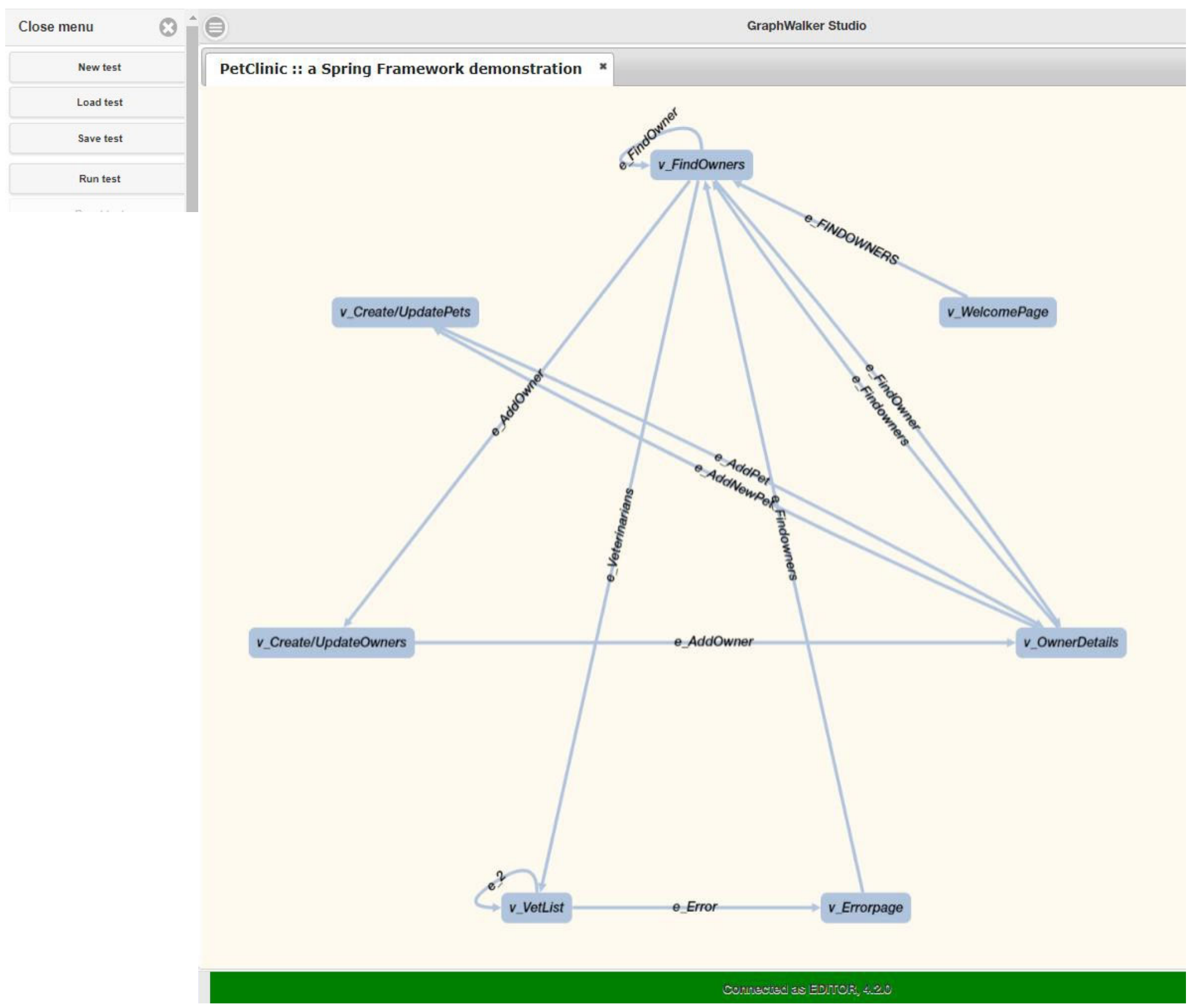

**Figure 7: Automatically-generated MBT model of the PetClinic web app, loaded and visualized in the GraphWalker tool**

### 5.4.2 SUT2: Task Manager web app

Using the same setup as the SUT1, we embedded the JavaScript logger and manually interacted with this SUT across different user roles, in the form of exploratory testing for about 20 minutes, ensuring that main paths are covered, e.g., task creation, editing, and status updating of tasks, in the Task Manager web app. We verified that the MBTModelGenerator tool successfully captured all the interactions precisely. Figure 8 shows the automatically-generated MBT model of this SUT by MBTModelGenerator, loaded and visualized in the GraphWalker tool.

We assessed and confirmed that the resulting MBT model reflected all functional branches, taken during the exploratory testing. Vertices include states such as: `v_Dashboard`, `v_AssignTask`, `v_EditTask`, and `v_VerifyCompletion`. Also, transitions (edges) such as `e_CreateTask`, `e_MarkComplete`, and `e_ViewAssignedTasks` were generated based on observed UI events.

When we pressed the *Run Test* button in GraphWalker, we observed that the automatically generated MBT model of this SUT too was executed successfully in the GraphWalker tool, without any issues. As a result, both empirical assessment criteria for this SUT were also passed: (1) Completeness and correctness of the generated MBT model; and (2) Successful loading and execution of the model in GraphWalker.

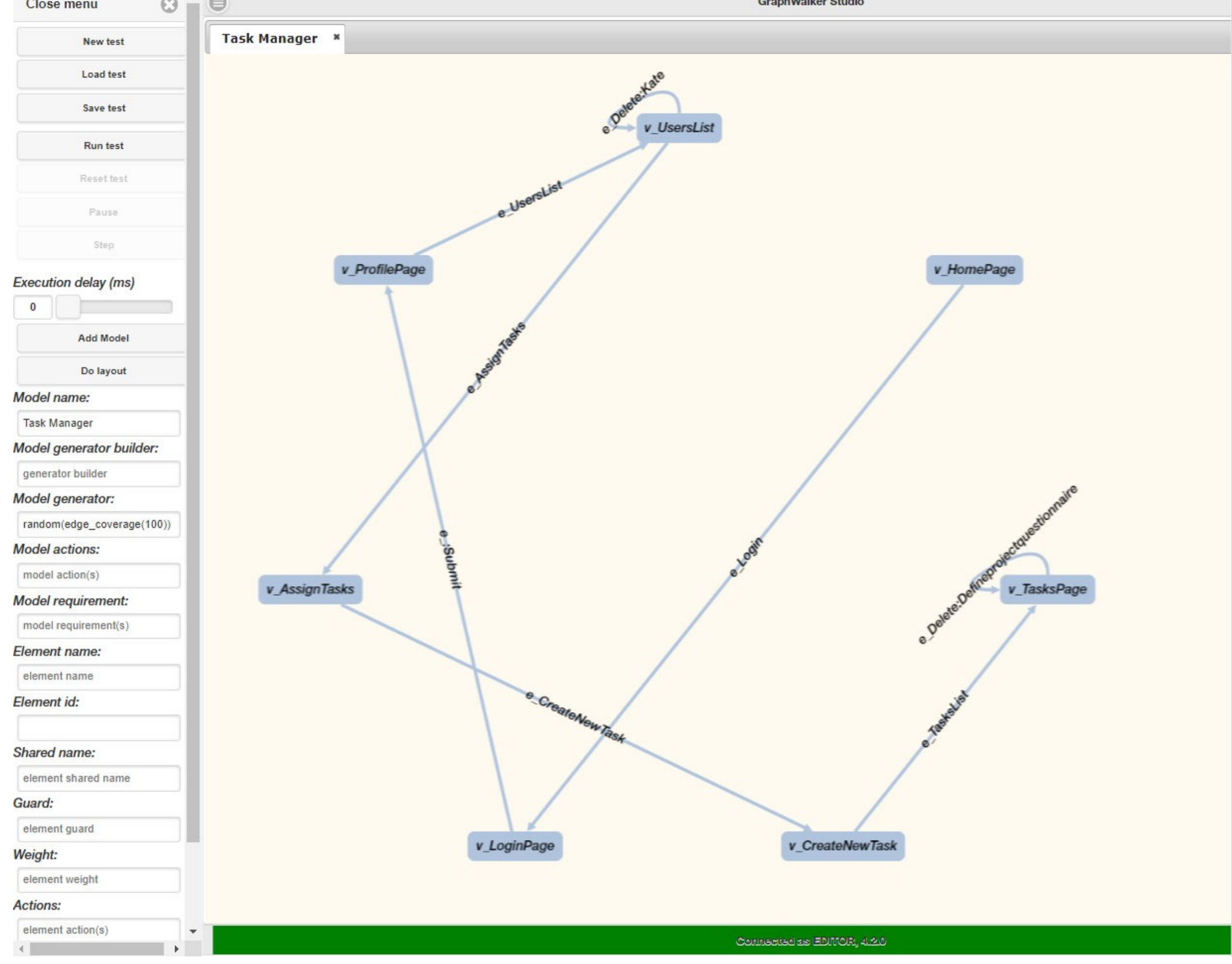

**Figure 8- Automatically-generated MBT model of the Task-Manager web app, loaded and visualized in the GraphWalker tool**

### 5.5 Estimation of effort (time) saving provided by MBTModelGenerator (RQ3)

To evaluate the time-saving potential of MBTModelGenerator (RQ3 of the empirical evaluation), we estimated and compared the effort required for the following two cases:

- Using our tool to automatically generate MBT models from clickstream data, followed by test execution; and
- Manually designing the MBT model and developing the corresponding Selenium Java code to implement edge transitions and assertions.

In our evaluation context, applying MBTModelGenerator—including exploratory testing of each SUT and automatic model generation—took approximately 5 minutes per application, i.e., the time taken for one-time exploratory testing. In contrast, based on practitioner experience at Testinium A.Ş., designing the MBT model manually and implementing test automation code typically requires at least 16 hours (two workdays) for a SUT of this size.

This corresponds to a time reduction of approximately 98%, offering a substantial decrease in manual effort. The efficiency gains are even more significant in real-world settings where clickstream data is already collected by usage analytics in server side. In such cases, the exploratory testing phase becomes unnecessary, and MBTModelGenerator can generate models fully offline, further reducing the required effort.

We shall note that, in more complex or larger-scale SUTs, to ensure correctness and completeness, the generated models should still be inspected by test engineers. Certain fixes and / or refinements may be necessary—such as resolving redundant edges, missing transitions, or semantically ambiguous states. Just like any software engineering tool, MBTModelGenerator is not "perfect," and further empirical validation through additional case studies is needed to further improve and generalize its applicability.

Nevertheless, even with such post-processing by test engineers, the combined required effort would remain substantially lower than designing MBT models and corresponding test scripts from scratch, affirming the tool's practical value in easing MBT adoption.

### 5.6 Observations and lessons learned

From the empirical evaluation conducted across both systems under test (SUTs), we summarize the following key observations and lessons:

- **Accuracy**: MBTModelGenerator effectively captured all major user navigation paths and UI state transitions observed during the exploratory testing sessions. As long as a flow was exercised at least once, it was accurately reflected in the generated MBT model. This demonstrates that the tool is suitable for generating behaviorally representative models from real usage data.
- **Robustness**: The tool correctly handled a wide range of frontend interactions, including form submissions, link and button clicks, and menu selections, without requiring any modifications to the SUT's internal structure or source code. This robustness ensures that the tool can be readily applied to a variety of modern web applications, regardless of their technology stack.
- **Simplicity and ease of integration**: One of the main practical benefits was the ease of integration. Only a lightweight JavaScript logger module needed to be included in the web application, which can often be done with minimal engineering effort. No instrumentation of the backend or changes to the SUT's architecture were required.
- **Completeness of the generated MBT models**:
    - MBTModelGenerator depends on actual user interaction traces (clickstream data). As such, any application behavior or path that was not exercised during the exploratory test session will not be represented in the resulting model. These cases require manual inspection and pruning by a test engineer to improve model conciseness and semantic clarity.
    - However, on the positive side, in many real-world contexts, large-scale offline clickstream data are already collected through analytics or monitoring systems. Statistically, the diverse behaviors of a large user base often result in coverage of nearly all navigable paths within a web application. In such cases, MBTModelGenerator can extract a more complete model that includes most—if not all—reachable nodes, transitions, and paths across a given SUT.
- **Need for human-in-the-loop**: While the tool offers a significant time-saving advantage, it is not intended to fully replace human test modeling expertise. In more complex systems, peer review and minor refinements of the generated models are still necessary to ensure completeness, eliminate noise (unwanted nodes or edges), and align the models with testing goals.

These findings confirm the practical viability of MBTModelGenerator as a lightweight, extensible, and automation-friendly solution for enabling MBT in web application testing environments. By reducing the manual effort traditionally associated with MBT model design and test code development, the tool facilitates broader industrial adoption of MBT, especially in settings where time and resources are constrained.

### 5.7 Limitations and threats to validity of the empirical study

While our results are promising, like any other empirical study, the above empirical evaluation is subject to several limitations:

- External validity: The evaluation was conducted on two open-source web applications. Although these are realistic and representative, broader validation across different domains, architectures, and project sizes is needed. Although the evaluation demonstrates that MBTModelGenerator produces correct and practically usable models for small-sized SUTs, a limitation of this study is that we did not include a medium or large-scale industrial SUT. Our choice of compact models (7–11 states / 8–12 transitions) was deliberate: they allowed us to validate correctness, traceability, and transformer behaviour in a controlled setting before introducing the additional confounding factors present in large models. In real projects, larger SUTs often require substantial post-processing—such as merging nodes, removing redundant transitions, and adding guard conditions—tasks that vary considerably across domains and teams. A rigorous measurement of the effort and time required for such refinement phases, or of how model maintenance evolves across successive SUT versions, was beyond the scope of this initial investigation. We therefore recognise that the generalisability of our results to complex systems remains an open question. Extending the evaluation to one or more medium- or large-scale SUTs, and systematically analysing the refinement workflow and its maintenance implications, constitutes an important direction for future work.

- Construct validity: The completeness of the generated models inevitably depends on the tester's exploratory behaviour and coverage during the capture phase. Any UI flow or feature not exercised during exploration will not be represented in the resulting model, potentially under-specifying the functional space of the SUT. This risk is inherent to all dynamic-capture–based MBT approaches. To mitigate this threat, our exploratory testing sessions—each approximately 20 minutes—were intentionally structured to traverse diverse UI paths and trigger as many state changes as possible. Nevertheless, we acknowledge that undiscovered or rarely used flows may still exist, and thus the generated models should be interpreted as *operational slices* of the SUT rather than complete behavioural specifications. Systematic exploration strategies or automated crawling techniques could be integrated in future work to further reduce this threat.
- Internal validity: Model correctness in this study was evaluated through manual inspection of the state/transition structure and by validating the models using GraphWalker, which successfully executed the generated paths without syntactic or runtime errors. This provides confidence that the produced models are structurally sound and semantically coherent with the navigated behaviour. However, we did not conduct formal test-effectiveness studies—e.g., fault seeding, mutation analysis, or comparative defect detection—to quantify how well the generated models would support MBT-based fault detection in industrial contexts. As such, the validity of the models is demonstrated for behavioural *fidelity* rather than for *testing efficacy*. Future research could complement our structural validation with empirical studies that measure fault-detection power across multiple SUT versions.
- Tool maturity: While MBTModelGenerator is stable and functional, it remains an evolving tool. Certain dynamic behaviours—particularly those associated with modern JavaScript frameworks, asynchronous DOM manipulation, or reactive UI patterns—can still present challenges for accurate state detection or may introduce noise that requires manual refinement. These limitations did not impair the small SUTs used in our study, but they may become more pronounced in large-scale or highly interactive systems. As the tool matures, improving its heuristics for state abstraction, event filtering, and dynamic UI handling will be essential to reduce manual post-processing effort and to enable more complex industrial evaluations.

## 6 DISCUSSIONS

This section offers broader insights beyond the evaluation, reported above. It highlights the tool's ongoing industrial use, discusses implications for MBT practice, and outlines the limitations and threats to validity of our empirical study.

### 6.1 Ongoing activities: Usage of the tool in industrial settings

Since its development, MBTModelGenerator has been actively used in internal testing activities at Testinium A.Ş. We note that this paper presented our evaluations using two open-source systems, and we chose not to report industrial case studies in this paper for three main reasons.

- **Confidentiality constraints**: The industrial SUTs in the company are client systems with proprietary test data that are subject to non-disclosure agreements. Releasing specific test models or findings would require significant anonymization or client consent, which was not feasible within the timeframe of this study.
- **Scope management**: Including industrial case studies would have required a broader study design and additional validation measures beyond the scope of this initial academic paper. We deliberately focused this paper on the technical design, tool implementation, and evaluation using two open-source systems, to keep the contributions well-scoped and reproducible.
- **Ongoing evaluation**: Industrial usage is still evolving. The tool is being incrementally refined based on feedback from internal teams, and more comprehensive industrial validation (including defect detection, ROI impact, and maintainability aspects) are being planned as of this writing (summer of 2025).

From the ongoing usage of the tool in actual industrial projects, we can share the following observations about the tool:

- The tool is actively applied for testing web-based test management, airline and e-commerce systems.
- It has helped bootstrap model-based test suites in projects lacking up-to-date documentation or formal specifications.
- Integration into internal pipelines has reduced the time required for initial model availability in regression testing efforts.

### 6.2 Implications for MBT practice in industry

As discussed earlier in Section 2.1, MBT adoption in industry has been hindered by the high effort and expertise required to manually design behavioral test models [12, 41-46]. The MBTModelGenerator initiative directly addresses this barrier by enabling semi-automated model derivation from actual user interaction data, reducing reliance on manual modeling.

This approach introduces several important implications for practice:

- It allows test teams to quickly generate baseline models even when application specifications or flowcharts are missing.
- It enables an incremental workflow where models are first auto-generated, then optionally refined by engineers.
- It aligns MBT practices with existing frontend instrumentation and logging infrastructures, allowing organizations to derive value from clickstream data already being collected for analytics or UX purposes.

By lowering the barrier to entry, this approach reframes MBT as a practical enhancement to existing QA practices, rather than a heavyweight process requiring specialized modeling expertise. In this context, the role of test engineers shifts from manually crafting models to inspecting and refining automatically derived ones, allowing greater focus on test objectives and quality assurance.

## 7 Conclusions and Future Work

This paper presented MBTModelGenerator, an open-source tool designed to ease the adoption of MBT in web application testing by automatically generating executable test models from clickstream data. The tool captures real user interactions through a lightweight JavaScript module and transforms them into GraphWalker-compatible MBT models. We described the design, implementation, and evaluation of the tool using two representative web applications.

Our evaluation, guided by the Goal-Question-Metric (GQM) framework [59], demonstrated that the tool can generate accurate and executable models, while providing a significant reduction—98%in the case study—in the effort typically required for model design and test-code development.

MBTModelGenerator has also been adopted internally at Testinium A.Ş., where it is being applied to real-world systems to automatically generate initial MBT models. The company has observed benefits in accelerating model availability, especially for legacy systems and undocumented applications. While we did not formally report those industrial case studies in this paper due to confidentiality constraints and scope management, broader industrial validations are ongoing and planned for future dissemination.

Looking forward, several future directions are planned to further improve and validate MBTModelGenerator:

- Model repair and noise reduction: Develop automated techniques to detect and prune redundant or low-value transitions, improving model clarity and test quality.
- AI-assisted model enrichment: Integrate AI techniques to suggest missing transitions, detect likely user paths, or auto-group states based on semantic similarity.
- Larger-scale validations: Conduct extensive case studies across diverse industrial systems, including large-scale applications at Testinium, to assess model completeness, maintainability, and test coverage.
- Support for dynamic web applications: Enhance the tool's capabilities to better handle dynamically generated DOM elements and single-page application (SPA) behaviors.
- Broader use of offline model generation: While MBTModelGenerator already supports generating models from stored clickstream logs, we plan to apply this feature more extensively in practice by integrating it with existing analytics pipelines and validating its effectiveness across diverse application domains.

These directions aim to increase the robustness and adaptability of MBTModelGenerator, enabling broader adoption of MBT practices across web-based software systems.

## Acknowledgements

Earlier support for this ongoing industry-academia collaborative project were provided by a TÜBİTAK-1509 grant from the Scientific and Technological Research Council of Turkey and a European ITEA4 R&D project named "TESTOMAT – The Next Level of Test Automation" (testomatproject.eu and itea4.org/project/testomatproject.html). The authors are grateful to the anonymous reviewers for their insightful comments for improving the quality of this paper.

## Declaration of Generative AI in the writing process

During the preparation of this work, the authors used the ChatGPT tools in order to improve language and readability. After using this tool, the authors reviewed and edited the content as needed and take full responsibility for the content of the publication.